\documentclass[a4paper,12pt]{article}
\pdfoutput=1
\usepackage{graphicx}
\usepackage{setspace}
\usepackage[title]{appendix}
\usepackage{inputenc}
\usepackage{ulem}
\usepackage{amsmath}
\usepackage{eurosym}
\usepackage{url}
\usepackage{authblk} 
\usepackage{epstopdf}
\usepackage{placeins}
\usepackage{tabularx}
\usepackage{longtable}
\usepackage{adjustbox}
\usepackage{booktabs}

\usepackage{bm}
\usepackage{float}
\usepackage[round]{natbib}
\usepackage{amsmath} 
\usepackage{amssymb} 
\usepackage{latexsym}
\usepackage{wasysym}
\usepackage{subfigure}
\usepackage[flushleft]{threeparttable}
\usepackage{color}
\usepackage{caption}

\begin{document}
	\title{Trade uncertainty
		impact on stock-bond correlations: Insights from conditional
		correlation models} 
	\date{}
	\author[$\dag$]{Demetrio Lacava}
	\author[$\ddagger$]{Edoardo Otranto\footnote{Corresponding author.}}
	
	\affil[$\dag$]{Department of Economics, University of Messina, Messina, via dei Verdi, 75, 98122, Italy (email: dlacava@unime.it)}
	\affil[$\ddagger$]{Department of Social Sciences and Economics, Sapienza University of Rome, Rome, Viale dell'Università, 36, 00185, Italy (email: edoardo.otranto@uniroma1.it)}
	
	\maketitle              

\begin{abstract} 
	
	This paper investigates the impact of Trade Policy Uncertainty (TPU) on stock–bond correlation dynamics in the United States. Using daily data on major U.S. stock indices and the 10-year Treasury bond from 2015 to 2025, we estimate correlation within a two-step GARCH-based framework, relying on multivariate specifications, including Constant Conditional Correlation (CCC), Smooth Transition Conditional Correlation (STCC), and Dynamic Conditional Correlation (DCC) models. We extend these frameworks by incorporating TPU index and a presidential dummy to capture effects of trade uncertainty and government cycles. The findings show that constant correlation models are strongly rejected in favor of time-varying specifications. Both STCC and DCC models confirm TPU’s central role in driving correlation dynamics, with significant differences across political regimes. DCC models augmented with TPU and political effects deliver the best in-sample fit and strongest forecasting performance, as measured by statistical and economic loss functions.\\
\end{abstract}
\vspace{2mm}

\section{Introduction}
Financial operators largely agree that asset allocation is one of the most critical decisions in the investment process. A key challenge in this context is that correlations between asset returns vary over time in response to economic shocks, policy changes, and geopolitical events. Accurately capturing and forecasting time-varying correlations is therefore crucial for effective for portfolio construction, diversification, and risk management. More in detail, diversification reduces exposure to idiosyncratic risk, but it depends directly on the correlation structure among assets. Assuming constant correlations can lead to suboptimal allocations as well as an underestimation of risk, particularly during periods of market turbulence or abrupt changes in investor risk perceptions. Moreover, asset prices often react in a similar way to common external shocks, suggesting that these shared influences should be explicitly taken into account when modeling asset comovement. For instance, \cite{Yilmazkuday:2025} documents a sharp decline in major U.S. stock indices associated with geopolitical risk \citep{Yilmazkuday:2024} and the COVID-19 pandemic \citep{Yilmazkuday:2023}. Economic and policy uncertainty (EPU) is also recognized as a key determinant of both oil–stock correlations \citep{Fang:Chen:Yu:Xiong:2018} and stock-bond correlations \citep{Connolly:Stivers:Sun:2005, Fang:Yu:Li:2017}. With respect to stock–bond correlations, evidence shows that they are positively related to inflation but negatively related to market uncertainty \citep{Andersson:Krylova:Vahamaa:2008}; in addition, they are also influenced by market liquidity \citep{Baele:Bekaert:Inghelbrecht:2010}, risk aversion \citep{Bekaert:Engstrom:Grenadier:2010}, and the U.S. presidential cycle \citep{Demirer:Gupta:2018}. 

More recently, the international political landscape has been dominated by pervasive trade uncertainty, with repeated tariff announcements and frequent adjustments to existing trade measures. From an investment viewpoint, understanding the impact of trade uncertainty on asset class correlations is particularly important, as sudden shifts in trade policy can materially alter diversification benefits and hedging effectiveness. To the best of our knowledge, however, the impact of trade uncertainty on stock–bond correlations has not yet been thoroughly investigated. Trade uncertainty can affect such correlations through multiple channels. First, higher trade uncertainty increases overall economic uncertainty and risk aversion, leading to the flight-to-safety phenomenon \citep{Pastor:Veronesi:2012, Brogaard:Detzel:2015}, in which investors reallocate wealth away from riskier assets toward safer ones, thereby generating negative stock–bond correlations. Second, trade shocks, such as the imposition of tariffs, may alter inflation expectations and, in turn, monetary policy decisions, producing periods of higher positive correlations between asset classes, as highlighted in the trade war context by \citet{Caldara:Iacoviello:Molligo:Prestipino:Raffo:2020}. Finally, trade uncertainty can dampen corporate earnings expectations and global aggregate demand. According to \citet{Bloom:2009}, this mechanism also contributes to form the flight-to-safety phenomenon and, as a consequence, to the time-varying nature of cross-asset correlations. All these mechanisms motivate the inclusion of Trade Policy Uncertainty (TPU) in multivariate volatility models, allowing us to capture an important source of time-varying correlations and to improve both in-sample fit and out-of-sample forecasts. \citet{Caldara:Iacoviello:Molligo:Prestipino:Raffo:2020} introduced a TPU index, which we employ in this study to assess whether TPU influences stock–bond correlations and whether such effects vary depending on the prevailing political agenda. 

From a computational perspective, modeling covariances and correlations raises two main challenges \citep{Bauwens:Laurent:Rombouts:2006}: (i) ensuring the positive definiteness of the estimated covariance (or correlation) matrix, and (ii) addressing the curse of dimensionality -- i.e., the exponential growth in the number of parameters as the number of assets increases -- which may render models computationally unfeasible. This issue is particularly severe in models such as the VECH GARCH \citep{Bollerslev:Engle:Wooldridge:1988} and the BEKK \citep{Engle:Kroner:1995}. An elegant and parsimonious solution, which accounts for much of the success of the approach, was proposed by \citet{Engle:2002}. His Dynamic Conditional Correlation (DCC) framework relies on a two-step estimation procedure: in the first step, univariate GARCH models are estimated for the conditional variances of each return series; in the second step, conditional correlations are estimated based on the standardized residuals. Different specifications have been provided for the representation of the conditional correlation matrix. Earlier contributions include \citet{Bollerslev:1990}, who proposed the Constant Conditional Correlation (CCC) model, in which correlations are fixed over time, while variances evolve according to univariate GARCH dynamics. 
\citet{Silvennoinen:Terasvirta:2015} later extended the CCC model by allowing smooth transitions between two correlation regimes, typically representing low and high correlation states, driven by an exogenous transition variable. The most widespread model in this framework, however, remains the DCC model of \citet{Engle:2002}, which introduces a GARCH-type structure for conditional correlations. 

In this paper, we extend these models along two dimensions: (i) by incorporating trade policy uncertainty, proxied by the TPU index, as a driver of correlation dynamics; and (ii) by accounting for the political cycle effect, captured through a dummy variable that reflects differences across broader policy regimes and macro-policy conditions over the political cycle. Our empirical analysis focuses on major U.S. stock indices -- namely the Standard and Poor's 500 (S\&P500), the Dow Jones Industrial Average, the Nasdaq Composite, and the Russell 2000 -- as representative of the stock market, and the 10-year Treasury Bond (T-Bond), as representative of the bond market, which is particularly suitable for investigating the structural (i.e., long-term) impact of trade uncertainty on correlations. 
Our results reveal several novel findings. First, stock–bond correlations alternate between positive and negative phases, with turning points coinciding with spikes in TPU. Second, Republican administrations are associated with stronger stock–bond comovements, thereby reducing the hedging role of bonds, while Democratic administrations tend to display weaker or negative correlations. Third, models that incorporate political regimes and TPU 
provide a superior representation of correlation dynamics compared with the baseline CCC. Finally, DCC models augmented with TPU and political effects deliver the best in-sample fit and significantly improve the accuracy of out-of-sample correlation forecasts, with direct implications for portfolio allocation and risk management.

The remainder of the paper is organized as follows. Section \ref{sec:visual_inspection} describes the data, while Section \ref{sec:theoretical_framework} outlines the model framework. Section \ref{sec:comparison} focuses  on the comparison of the models: subsection \ref{sec:insample} reports the in-sample analysis, whereas subsection \ref{sec:forecast} provides the out-of-sample evaluation. Section \ref{sec:conclusion} concludes with some final remarks.

\section{Tracking the Joint Evolution of Correlations and Trade Uncertainty}
\label{sec:visual_inspection}
To examine the impact of trade uncertainty on stock-bond correlations, we rely on the TPU index\footnote{Data are available at https://www.matteoiacoviello.com/tpu.htm.} by \citet{Caldara:Iacoviello:Molligo:Prestipino:Raffo:2020}. Similar to the EPU index of \cite{Baker:Bloom:Davis:2016}, the TPU index measures the frequency of joint occurrences of terms related to trade policy and uncertainty across major newspapers. 
The sample spans the period from January 5, 2015 to July 18, 2025, covering four different U.S. government administrations.\footnote{They refer to Obama's second term, Trump's first, Biden's, Trump's second.} This enables us to analyze how shifts in political agendas (e.g., in terms of fiscal stance, trade policy orientation, and the international policy environment) may shape policy uncertainty and, in turn, influence  stock-bond correlation dynamics. For the same period, we collect daily data (sourced from Yahoo Finance) for the S\&P 500, Dow Jones Industrial Average, Nasdaq Composite, and Russell 2000 as proxies for the equity market, and the 10-year U.S. Treasury bond (T-Bond) as a representative of the bond market. The choice of the 10-year maturity allows us to investigate the structural (i.e., long-term) impact of trade uncertainty on correlations. In contrast, a shorter-term instrument such as the 3-month Treasury Bill (T-Bill) would mainly capture short-term market fluctuations. Since trade uncertainty primarily arises from tariff policies, whose effects are expected to persist over time, the T-Bond is a more suitable proxy.
\begin{figure}[t]
	\centering
	{\includegraphics[height=9cm,width=14cm]{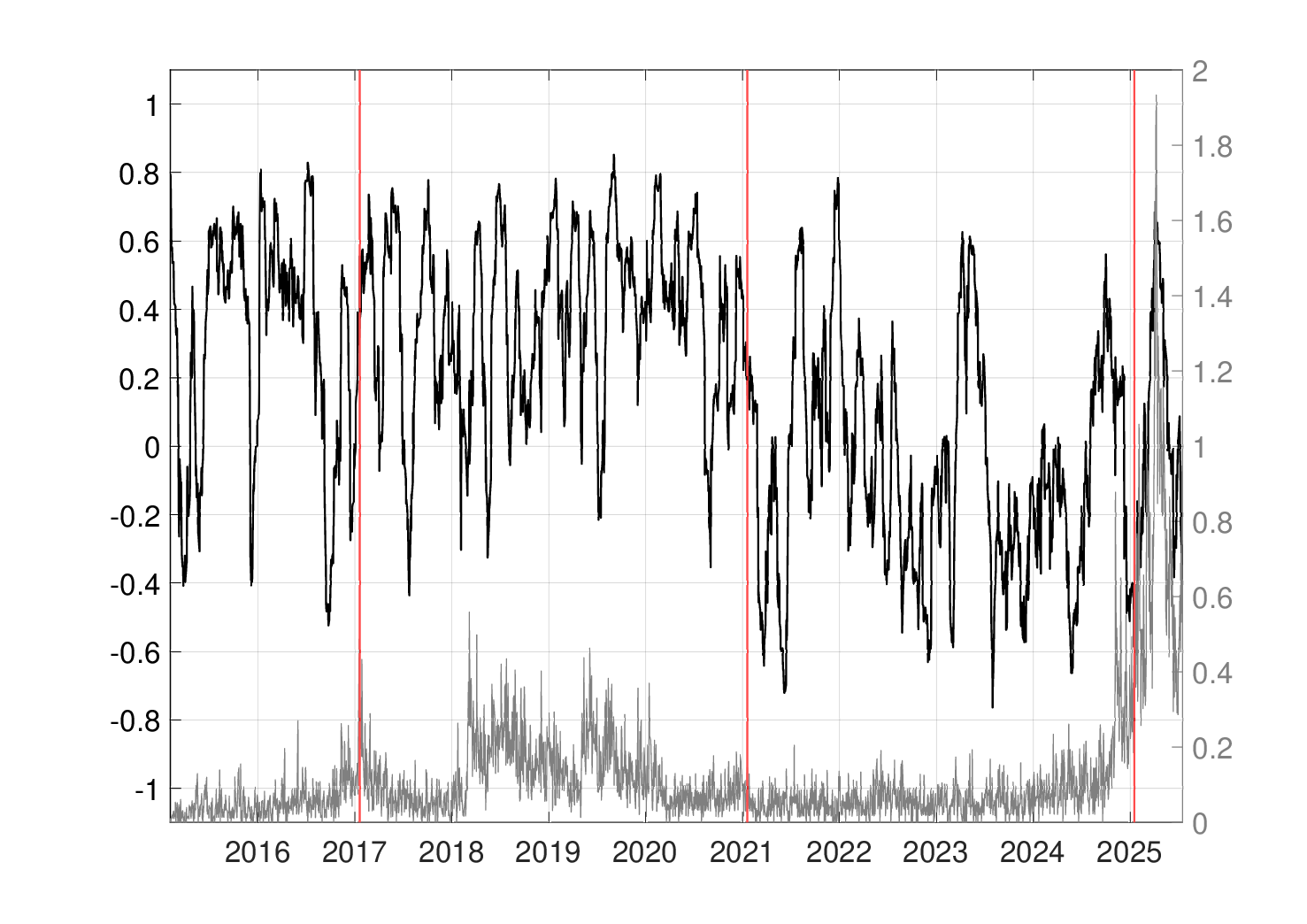}}
	\caption{Monthly rolling T-Bond-S\&P500 correlation (black line, left axis), Trade Policy Uncertainty (TPU) index (gray line, right axis), and Inauguration Day (vertical red line). Sample period: January 5, 2015 -- July 18, 2025.}
	\label{fig:rollingcorr_tpu}
\end{figure}

Figure \ref{fig:rollingcorr_tpu} illustrates the joint evolution of the monthly (22 terms) rolling correlations between the T-bond and the S\&P500 (black line, left axis) and the TPU index (gray line, right axis), with vertical red lines marking each U.S. presidential inauguration  within the sample period. A clear pattern emerges, characterized by alternating phases of positive and negative correlations. Notably, turning points from negative to positive correlations frequently coincide with spikes in TPU.
The rolling correlation remains relatively stable until 2018, encompassing the final phase of the Obama administration and the early part of Trump’s first term. A pronounced rise in correlation ($+159\%$) occurs following the imposition of tariffs on March 21, 2018 (25\% on imported steel and 10\% on imported aluminum). Similarly,  correlations peaked at around 0.6 in the days immediately after the May 10, 2019 announcement of additional tariffs against China. Clustering is also evident in the TPU index: a positive trend in TPU (and in correlations) reappears in 2025, while the lower and more stable TPU levels observed during 2021–2025 suggest that the prevailing policy agenda may influence trade policy uncertainty, which in turn affects stock–bond comovements. Overall, these observations provide strong motivation for employing conditional correlation models that explicitly incorporate both political regime changes and trade policy uncertainty.

\begin{table}[h!]
	\caption{Pairwise sample correlations and average TPU for the full sample and selected subsamples.  Government administrations are indicated in parentheses. Sample period: January 5, 2015 -- July 18, 2025.}\label{tab:sample_corr}
\begin{adjustbox}{max width=0.7\linewidth,center}
		\begin{tabular}{lcccc}
			\hline
			\multicolumn{5}{c}{January   05, 2015 -- April 30, 2025 }   \\[1.5mm]
			& S\&P 500 & Dow Jones & Nasdaq & Russell 2000 \\
			\hline
			Dow Jones   & 0.954    &           &        &             \\
			Nasdaq      & 0.950    & 0.841     &        &             \\
			Russel 2000 & 0.871    & 0.853     & 0.831  &             \\
			T-Bond      & 0.162    & 0.189     & 0.120  & 0.139       \\
			TPU average & \multicolumn{4}{c}{0.121}                   \\[2mm]
			\hline
			\multicolumn{5}{c}{January 05, 2015 --   January 19, 2017 (Obama 2)} \\[1.5mm]
			& S\&P 500 & Dow Jones & Nasdaq & Russell 2000 \\
			Dow Jones   & 0.977    &           &        &             \\
			Nasdaq      & 0.951    & 0.901     &        &             \\
			Russel 2000 & 0.874    & 0.838     & 0.883  &             \\
			T-Bond      & 0.357    & 0.369     & 0.344  & 0.376       \\
			TPU average & \multicolumn{4}{c}{0.051}                   \\[2mm]
			\hline
			\multicolumn{5}{c}{January 20, 2017 --   January 19, 2021 (Trump 1)} \\[1.5mm]
			& S\&P 500 & Dow Jones & Nasdaq & Russell 2000 \\
			Dow Jones   & 0.976    &           &        &             \\
			Nasdaq      & 0.952    & 0.889     &        &             \\
			Russel 2000 & 0.896    & 0.891     & 0.843  &             \\
			T-Bond      & 0.389    & 0.395     & 0.332  & 0.371       \\
			TPU average & \multicolumn{4}{c}{0.126}                   \\[2mm]
			\hline
			\multicolumn{5}{c}{January 20, 2021 --   January 17, 2025 (Biden)} \\[1.5mm]
			& S\&P 500 & Dow Jones & Nasdaq & Russell 2000 \\
			Dow Jones   & 0.917    &           &        &             \\
			Nasdaq      & 0.951    & 0.780     &        &             \\
			Russel 2000 & 0.831    & 0.810     & 0.794  &             \\
			T-Bond      & -0.091   & -0.053    & -0.109 & -0.104      \\
			TPU average & \multicolumn{4}{c}{0.079}                   \\[2mm]
			\hline
			\multicolumn{5}{c}{January 21, 2025 --   July 18, 2025 (Trump 2)}   \\[1.5mm]
			& S\&P 500 & Dow Jones & Nasdaq & Russell 2000 \\
			Dow Jones   & 0.949    &           &        &             \\
			Nasdaq      & 0.977    & 0.878     &        &             \\
			Russel 2000 & 0.926    & 0.899     & 0.899  &             \\
			T-Bond      & 0.254    & 0.169     & 0.304  & 0.200       \\
			TPU average & \multicolumn{4}{c}{0.715}                  \\
			\bottomrule
		\end{tabular}
\end{adjustbox}
\end{table}

Table \ref{tab:sample_corr} reports pairwise sample correlations among the selected indices and the average value of the TPU index, for both the full sample and selected sub-periods. Correlations among equity indices remain stable across sub-samples, while those involving the 10-year T-bond are more variable. From a portfolio diversification perspective, low or negative stock-bond correlations are desirable, as losses in one asset class can be offset by gains in the other. Across the full sample, correlations are close to zero, ranging from 0.120 (with the Nasdaq) to 0.189 (with the Dow Jones).
During President Obama’s second term, stock–bond correlations increased relative to the full sample, reaching up to 0.38, despite low levels of both policy and trade uncertainty. Correlations rose further, particularly for the S\&P 500 and the Dow Jones, during Trump’s first term, when tariffs on solar panels and aluminum were introduced, but turned negative during the Biden administration. Since both administrations implemented tariffs, this pattern suggests that the dynamics of stock-bond correlations are more strongly linked to the level of uncertainty than to the implementation of tariffs themselves. This interpretation is consistent with the evolution of the TPU index: relative to the Obama administration, trade policy uncertainty increased by roughly 146\% during Trump's first term (2017–2021), but declined by 38\% during Biden administration (2021–2025). Exceptionally high levels of uncertainty are recorded in 2025, accompanied by a renewed increase in stock–bond correlations -- up to 0.200 (Russell 2000) and 0.304 (Nasdaq) -- with an average TPU index of 0.715. 
Taken together, these findings suggest a breakdown of the traditional diversification benefits of bonds during periods of elevated uncertainty, when bonds may share a similar level of perceived risk as stocks. 


\section{Conditional Correlation Models}
\label{sec:theoretical_framework}

All conditional correlation models considered here share a common two-step estimation procedure \citep{Engle:2002}. In the first step, conditional variances are estimated; in the second, conditional correlations are obtained.

Let $\bm{r}_t$ be an ($N \times 1$) vector 
of daily returns,  with conditional covariance matrix $\mathbf{H}_t$. We assume 
$$\bm{r}_t| F_{t-1} \sim N(0, \mathbf{H_t}), \qquad t = 1, \ldots , T$$ 
and
$$\mathbf{H_t} = \mathbf{S_t R_t S_t},$$
where $F_{t-1}$ denotes the information set, $\mathbf{S}_t$ is a diagonal matrix of conditional standard deviations, and $\mathbf{R}_t$ is the conditional correlation matrix.

For each asset \textit{i}, conditional standard deviations are obtained from a univariate GJR-GARCH model \citep{Glosten:Jaganathan:Runkle:1993}, where the conditional variance $h^2_{i,t}$ (the $i$-th diagonal element of $\mathbf{H_t}$)  is defined as
\begin{equation}
	h^2_{i,t}=\omega_i+\alpha_i r^2_{i,t-1}+\beta_i h^2_{i,t-1} +\gamma_i r^2_{i,t-1} I_{i,t-1}, \qquad i=1,\dots,N \label{eq:gjr_garch}
\end{equation}
where $I_{i,t}$ is an indicator function taking the value of 1 when the $i-$th element of the vector of returns, $r_{i,t}$, is negative and 0 otherwise. Here $ r^2_{i,t-1}$ captures recent shocks, and $h^2_{i,t-1}$ reflects the impact of the lagged conditional volatility. Stationarity requires $(\alpha_i+\beta_i+\frac{\gamma_i}{2})<1$, while $\omega_i>0,\alpha_i\ge 0,\beta_i\ge 0,\gamma_i\ge 0$ ensure positivity.

De-garched returns are computed as $\tilde{\epsilon}_{i,t} = r_{i,t}/h_{i,t}$, and the second step estimate provides $\mathbf{R}_t$,  using specifications that possibly allow for exogenous regressors.

Under conditional normality, correlations are estimated by Maximum Likelihood, with log-likelihood 
\begin{equation}
	L(\bm{\theta}) = -\frac{TN}{2}log(2\pi) -\frac{T}{2} \log\left( |\mathbf{R_t}| \right) 
	-\frac{1}{2} \sum_{t=1}^T 
	\tilde{\bm{\epsilon}}_{t,\cdot} \, 
	\mathbf{R_t}^{-1} \, 
	\tilde{\bm{\epsilon}}_{t,\cdot}^\prime, \label{eq:loglik}
\end{equation}
where $\bm{\theta}$ denotes the parameter vector and $\tilde{\bm{\epsilon}}_{t,\cdot}$ the $t$-th row of the matrix of de-garched returns. Robust standard errors \citep{White:1980} are used to account for potential heteroskedasticity.

We now describe the conditional correlation specifications adopted and the main empirical findings. The dataset used for the in-sample analysis ends on February 24, 2023, while the remaining observations will be considered for the out-of-sample analysis developed in subsection \ref{sec:forecast}.

\subsection{Constant Conditional Correlation Models}
\label{sec:ccc}

As a benchmark, we first consider the Constant Conditional Correlation (CCC) model of \citet{Bollerslev:1990}, where the correlation matrix $\mathbf{R}$ is assumed to be time-invariant. Covariances remain time-varying since they scale with the conditional volatilities from the first-step GARCH estimates. Positive definiteness of $\mathbf{H}_t$ is guaranteed under the same conditions as in \citet{Engle:Kroner:1995}.
To estimate $\mathbf{R}_t = \mathbf{R}$, \cite{Bollerslev:1990} suggests using the sample correlation of de-garched returns. Given the manageable size of the correlation matrix ($5 \times 5$), which implies the estimation of 10 parameters, we prefer to maximize the log-likelihood in Eq.(\ref{eq:loglik}), using a triangular decomposition that guarantees positive semidefiniteness. Denoting $\rho_{ij}$ the generic element of the matrix $\mathbf{R}$, we consider the lower triangular matrix $\bm{P}=\{p_{ij}\}$ with elements:
\[
\begin{array}{ll}
	p_{11}=1&\\
	p_{ij}=\rho_{ij} & i=2,\dots,5; j=1,\dots,i-1 \\
	p_{ii}=\left(1-\sum_{j=1}^{i-1}p_{ij}^2\right)^{1/2} & i=2,\dots,5\\
\end{array}
\]
Thus, we obtain a positive semidefinite correlation matrix from $\bm{R}=\bm{P}\bm{P}'$.

Table \ref{tab:results_ccc} reports the estimated correlations with robust standard errors. All correlations are statistically significant at the 1\% level. As expected, market indices are highly interconnected, with correlations ranging from 0.905 (Dow Jones–Russell 2000) to 0.972 ($S\&P500$–Dow Jones). Correlations with the 10-year T-bond  are lower, but still substantial (average 0.548), peaking at 0.578 with the Dow Jones. Compared to raw sample correlations, which are roughly one-third of these values, the CCC model tends to overestimate stock–bond linkages.
\begin{table}[t]
	\caption{Estimated correlations matrix from the CCC model with robust standard errors \citep{White:1980} in parenthesis. Sample period: January 5, 2015 -- February 24, 2023.}\label{tab:results_ccc}
	\begin{adjustbox}{max width=0.7\linewidth,center}
		\begin{tabular}{lcccc}
			\hline
			\multicolumn{1}{c}{}          & $S\&P500$ & Dow Jones  & Nasdaq & Russell 2000 \\
			\hline
			Dow Jones     & 0.9720    &       &        &       \\
			& (0.0013) &       &        &\\
			Nasdaq    & 0.9680    & 0.9047 &      &       \\
			& (0.0015)  & (0.0047) &        &\\
			Russell 2000     & 0.9270    & 0.9073 & 0.9096 &     \\
			& (0.0033)&(0.0036) & (0.0050) &\\
			$T-Bond$  & 0.5436    & 0.5783 & 0.5052 & 0.5650\\       
			&(0.0297) & (0.0266) &(0.0318)&(0.0289)\\
			\bottomrule
		\end{tabular}
	\end{adjustbox}
\end{table}
Finally, under the assumption of constant correlations, the cross product of the de-garched returns, $\tilde{\epsilon}_{i,t}\tilde{\epsilon}_{j,t}$ ($i \neq j$), should exhibit no serial correlation.
However, Ljung–Box test results (Table \ref{tab:lb_res_bond}) reject the null at the 5\% level for all bond-related pairs, pointing to serial dependence (critical values of the corresponding chi-squared distribution are reported in the last row of the table). This provides evidence in favor of a time-varying correlation specification.

\begin{table}[t]
	\caption{Ljung-Box test statistics for the cross product of the de-garched returns with the corresponding critical value (last row). Sample period: January 5, 2015 -- February 24, 2023.}\label{tab:lb_res_bond}
	\begin{adjustbox}{max width=1\linewidth,center}
		\begin{tabular}{l|ccccccccccccc}
			\hline
			
			\multicolumn{1}{c}{$T-Bond$ vs}            & Lag 1   & Lag 2   & Lag 3   & Lag 4   & Lag 5   & Lag 6   & Lag 7   & Lag 8   & Lag 9   & Lag 10  & Lag 15  & Lag 20   \\
			\hline
			$S\&P500$            & 17.7164 & 30.9564 & 33.1594 & 44.9308 & 50.9474 & 56.4707 & 60.1150 & 63.4548 & 63.9716 & 65.4095 & 90.0262 & 110.2544 \\
			Dow Jones              & 15.0543 & 29.0535 & 30.1883 & 41.8850 & 45.6699 & 53.2492 & 54.5826 & 56.9486 & 57.2172 & 60.8605 & 81.3625 & 95.0830  \\
			Nasdaq               & 14.9247 & 23.4242 & 25.9365 & 34.8321 & 42.1830 & 45.7476 & 51.6326 & 55.2625 & 55.6850 & 56.6138 & 85.0104 & 102.5510 \\
			Russell 2000              & 18.1042 & 35.2715 & 35.8245 & 43.0488 & 58.0648 & 63.8264 & 67.2017 & 68.8044 & 69.4328 & 72.3726 & 99.4261 & 121.8535 \\
			$\chi^2_{0.05,lags}$ & 3.8415  & 5.9915  & 7.8147  & 9.4877  & 11.0705 & 12.5916 & 14.0671 & 15.5073 & 16.9190 & 18.3070 & 24.9958 & 31.4104 
			\\       
			\bottomrule
		\end{tabular}
	\end{adjustbox}
\end{table}


Basing on the discussion in Section \ref{sec:visual_inspection}, as a first extension, we allow correlations to vary with the political cycle. Specifically, we introduce a dummy variable $D_t$ equal to 1 during Republican administrations and 0 otherwise. The resulting CCC with Political Effect (CCC-PE) model is based on two correlation matrices:

\begin{equation}
	\mathbf{R}_t=\mathbf{R}_1D_t+\mathbf{R}_2(1-D_t),  \label{eq:ccc-pe} 
\end{equation}
where $\mathbf{R}_1$ applies under Republican administrations and $\mathbf{R}_2$ under Democratic ones.

Table \ref{tab:results_double_ccc} reports the estimation results. Stock market correlations remain above 0.9 in both regimes, showing little sensitivity to the political cycle. By contrast, bond–stock correlations display clear variation. During Republican administrations, they average 0.707, with a peak of 0.733 against the Russell 2000. Under Democratic administrations, the average falls to 0.367, ranging from 0.325 (vs. Nasdaq) to 0.409 (vs. Dow Jones). Thus, while stock–stock correlations change by only 0.001, bond–stock correlations are markedly lower under Democratic governments -- by 77\% to 104\% compared to Republican periods.
This asymmetry may reflect structural differences in fiscal and monetary policies. Historically, Democratic administrations are associated with lower T-Bond yields relative to equity returns, enhancing diversification benefits. Conversely, higher bond–stock correlations during Republican periods may indicate stronger transmission of macroeconomic shocks -- such as tariff escalations -- across asset classes, reducing the hedging role of government bonds.

\begin{table}[t]
	\caption{Estimated correlations matrix from the CCC-PE model with robust standard errors \citep{White:1980} in parenthesis and the corresponding administration. Sample period: January 5, 2015 -- February 24, 2023.}\label{tab:results_double_ccc}
	\begin{adjustbox}{max width=0.7\linewidth,center}
		\begin{tabular}{lcccc}
			\hline
			\multicolumn{1}{c}{}          & $S\&P500$ & Dow Jones  & Nasdaq & Russell 2000 \\
			\hline
			\multicolumn{5}{c}{$\mathbf{R}_1$ -- (Republican administration)}\\
			Dow Jones     & 0.9741    &       &        &        \\
			&(0.0016)&       &        &        \\
			Nasdaq    & 0.9682    & 0.9093 &       &        \\
			&(0.0024)&(0.0066)&        &        \\
			Russell 2000     & 0.9261    & 0.9107 & 0.8956 &       \\
			&(0.0048)&(0.0048)&(0.0090)&        \\
			$T-Bond$  & 0.7075    & 0.7230 & 0.6632 & 0.7333\\
			&(0.0203)&(0.0183)&(0.0231)&(0.0155)\\[2mm]
			\hline
			\multicolumn{5}{c}{$\mathbf{R}_2$ -- (Democratic administration)}\\
			Dow Jones    & 0.9695    &      &        &        \\
			&(0.0022)&       &        &        \\
			Nasdaq   & 0.9673    & 0.8985 &      &        \\
			&(0.0018)&(0.0069)&        &        \\
			Russell 2000    & 0.9281    & 0.9032 & 0.9232 &      \\
			&(0.0046)&(0.0058)&(0.0054)&        \\
			$T-Bond$ & 0.3599    & 0.4094 & 0.3250 & 0.3744\\
			&(0.0683)&(0.0638)&(0.0719)&(0.0711)\\
			\bottomrule
		\end{tabular}
	\end{adjustbox}
\end{table}

\subsection{Smooth Transition Conditional Correlation Models}
\label{sec:stcc}
A more flexible specification is the Smooth Transition Conditional Correlation (STCC) model of \citet{Silvennoinen:Terasvirta:2015}, where correlations evolve according to a smooth transition function. The correlation matrix is given by
\begin{equation}
	\mathbf{R}_t=\mathbf{R}_1G_t+\mathbf{R}_2(1-G_t) \label{eq:stcc}
\end{equation}
where $G_t$ is a logistic (smooth transition) function of the transition variable $x_t$ (here the TPU index); we call this model STCC with Trade Uncertainty Effect (STCC-TUE):
\begin{equation}
	G_t=(1+\exp^{-\varphi(x_{t-1}-c)})^{-1}. \label{eq:st_function}
\end{equation}
Here, $c$  is the location parameter, which defines the  threshold of the transition,  and the slope parameter $\varphi > 0$ controls the speed of transition, which becomes abrupt as $\varphi \to \infty$ \citep[see][for further details about the theoretical properties of the model]{Silvennoinen:Terasvirta:2015}.

Table \ref{tab:results_stcc} reports the estimation results. The slope parameter $\varphi$  is highly significant at the 1\% level, confirming that correlations increase with the TPU index. 
In $\mathbf{R}_1$, stock market correlations are high, with the maximum between $S\&P500$ and Nasdaq (0.984), and bond–stock correlations are also strong, ranging from 0.822 (T-Bond–Nasdaq) to 0.861 (T-Bond–Russell 2000). This indicates stronger integration under high TPU, particularly between T-Bonds and small-cap equities, which are more exposed to trade uncertainty.
In contrast, in $\mathbf{R}_2$ the T-bond correlations are negative and statistically not significant, pointing to weak integration when TPU is low. Since the location parameter ${c}$ is not significant and TPU exceeds zero in 95\% of the sample, state 1 can be considered the prevailing regime.

\begin{table}[t]
	\caption{Estimated correlations matrix from the STCC-TUE model with robust standard errors \citep{White:1980} in parentheses. Sample period: January 5, 2015 --- February 24, 2023.}\label{tab:results_stcc}
	\begin{adjustbox}{max width=0.7\linewidth,center}
		\begin{tabular}{lcccc}
			\hline
			\multicolumn{1}{c}{}          & $S\&P500$ & Dow Jones  & Nasdaq & Russell 2000 \\
			\hline
			\multicolumn{5}{c}{$\mathbf{R}_1$}\\
			Dow Jones    & 0.9778 &        &        &        \\
			& (0.0032) &        &        &        \\
			Nasdaq       & 0.9835 & 0.9442 &        &        \\
			& (0.0043)& (0.0180) &        &        \\
			Russell 2000 & 0.9198 & 0.8785 & 0.9406 &        \\
			& (0.0156) &(0.0134) & (0.0200)  &        \\
			$T-Bond$       & 0.8371 & 0.8291 & 0.8221 & 0.8611\\
			&  (0.0533)& (0.0398)&(0.0638) & (0.0566) \\[2mm]
			\hline
			\multicolumn{5}{c}{$\mathbf{R}_2$}\\
			Dow Jones    & 0.9567  &         &         &         \\
			& (0.0088) &        &        &        \\
			Nasdaq       & 0.9282  & 0.8022  &         &         \\
			& (0.0181) & (0.0452) &        &        \\
			Russell 2000 & 0.9448  & 0.9790  & 0.8309  &         \\
			& (0.0371) & (0.0370) & (0.0552) &        \\
			$T-Bond$       & -0.2021 & -0.0612 & -0.3041 & -0.1853\\
			& (0.2402) & (0.2282) & (0.2321) & (0.2266) \\
			\hline
			\multicolumn{1}{c}{}&$\varphi$ &  6.7608 & $c$     &-0.0559 \\ 
			\multicolumn{1}{c}{}&  & (1.6981) & &(0.0729) \\ 
			\bottomrule
		\end{tabular}
	\end{adjustbox}
\end{table}

\begin{figure}[t]
	\centering
	\subfigure{\includegraphics[height=9cm,width=14cm]{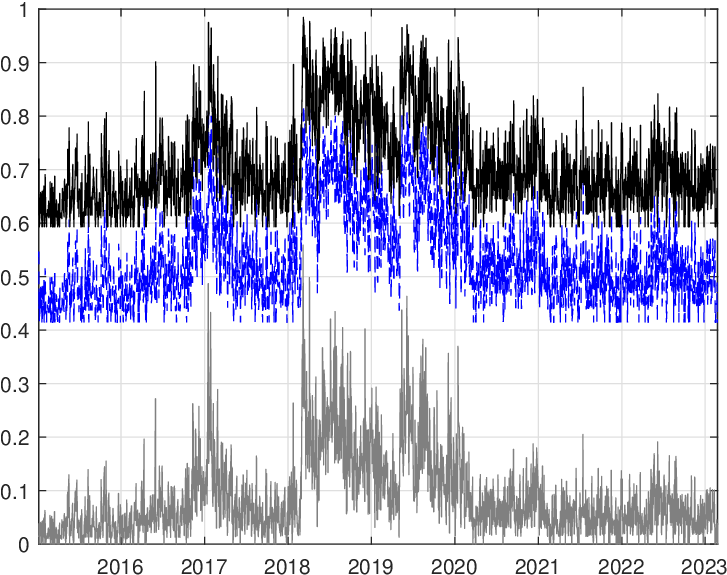}}
	\caption{ST-CC Estimated  $S\&P500$-$T-Bond$ correlation (dotted-blue line), smooth transition function (black line), and Trade Policy Uncertainty (TPU) index (dashed-gray line).  Sample period: January 5, 2015 -- February 24, 2023.}
	\label{fig:est_stcc}
\end{figure}

Figure \ref{fig:est_stcc} illustrates  the correlation between T-Bond and $S\&P500$ (dotted-blue line), along with the smooth transition function $G_t$ (black line) and the TPU index (dashed-gray line). The correlation rises above 0.8 during TPU peaks, while it falls below 0.6 in periods of persistently low TPU, mirroring the dynamics of the transition function.

Including the dummy variable $D_t$ to distinguish Republican from Democratic administrations allows the STCC model to capture different stock–bond correlation dynamics. We specify the following model:

\begin{align}
	\mathbf{R}_t = &\left[ \mathbf{R}_1  G_t^{(1)} + \mathbf{R}_2  (1 - G_t^{(1)}) \right] D_t + \left[ \mathbf{R}_3G_t^{(2)} + \mathbf{R}_4(1 - G_t^{(2)}) \right] (1-D_t)  \notag \\[2mm]
	G_t^{(d)} =  & (1+\exp^{-\varphi_d(x_{t-1}-c_d)})^{-1} \quad d=1,2 \label{eq:stcc_tupe}  
\end{align}
Here, $\mathbf{R}_1$ and $\mathbf{R}_2$ are the  correlation matrices corresponding to the Republican terms, while $\mathbf{R}_3$ and $\mathbf{R}_4$ apply to the Democratic periods. The smooth transition functions allow different location ($c_d$) and slope ($\varphi_d$) parameters across regimes $d=1,2$, representing the Republican and Democratic administrations, respectively. We refer to this specification as the STCC with Trade Uncertainty and Political Effect (STCC-TUPE).
This specification is similar to that of the Double-STCC model proposed by \citet{Silvennoinen:Terasvirta:2009}, with the important difference that in STCC-TUPE the changes in the STCC specification are deterministic, based on $D_t$, while in Double-STCC ${\bm R}_t$ is a convex combination of four positive definite matrices.

\begin{table}[t]
	\caption{Estimated correlations matrix from the STCC-TUPE model with robust standard errors \citep{White:1980} in parentheses and the corresponding administration. Sample period: January 5, 2015 -- February 24, 2023.}\label{tab:results_double_stcc}
	\begin{adjustbox}{max width=1\linewidth,center}
		\begin{tabular}{lccccccccc}
			\hline
			\multicolumn{1}{c}{}          & $S\&P500$ & Dow Jones  & Nasdaq & Russell 2000 & & $S\&P500$ & Dow Jones  & Nasdaq & Russell 2000 \\
			\hline
			\multicolumn{5}{c}{Republican administration} & \multicolumn{5}{c}{Democratic administration} \\
			\multicolumn{5}{c}{$\mathbf{R}_1$} &    \multicolumn{5}{c}{$\mathbf{R}_3$}\\
			Dow Jones    & 0.9769    &           &          &               & Dow Jones        & 0.7290    &           &          &              \\
			& (0.0021)  &           &          &               &                  & (0.4912)  &           &          &              \\
			Nasdaq       & 0.9795    & 0.9349    &          &               & Nasdaq           & 0.5460    & -0.0525   &          &              \\
			& (0.0029)  & (0.0069)  &          &               &                  & (0.8868)  & (0.5766)  &          &              \\
			Russell 2000 & 0.9408    & 0.9026    & 0.9499   &               & Russell 2000     & 0.8802    & 0.7504    & 0.4824   &              \\
			& (0.0058)  & (0.0097)  & (0.0113) &               &                  & (0.7075)  & (0.5309)  & (0.8445) &              \\
			$T-Bond$     & 0.7384    & 0.7432    & 0.7169   & 0.7501        & $T-Bond$         & 0.0491    & 0.0479    & -0.0225  & -0.0083      \\
			& (0.0304)  & (0.0280)  & (0.0324) & (0.0257)      &                  & (0.0756)  & (0.0415)  & (0.2111) & (0.2255)     \\[2mm]
			\hline
			\multicolumn{5}{c}{Republican administration} & \multicolumn{5}{c}{Democratic administration}\\
			
			\multicolumn{5}{c}{$\mathbf{R}_2$} &    \multicolumn{5}{c}{$\mathbf{R}_4$}\\
			Dow Jones    & 0.9596    &           &          &               & Dow Jones        & 0.9695    &           &          &              \\
			& (0.0150)  &           &          &               &                  & (0.0022)  &           &          &              \\
			Nasdaq       & 0.9117    & 0.7789    &          &               & Nasdaq           & 0.9673    & 0.8985    &          &              \\
			& (0.0292)  & (0.0789)  &          &               &                  & (0.0018)  & (0.0068)  &          &              \\
			Russell 2000 & 0.8536    & 0.9491    & 0.6273   &               & Russell 2000     & 0.9281    & 0.9032    & 0.9232   &              \\
			& (0.0508)  & (0.0281)  & (0.1287) &               &                  & (0.0046)  & (0.0049)  & (0.0042) &              \\
			$T-Bond$     & 0.5548    & 0.6235    & 0.3917   & 0.6544        & $T-Bond$         & 0.3599    & 0.4095    & 0.3250   & 0.3744       \\
			& (0.1600)  & (0.1358)  & (0.2035) & (0.1172)      &                  & (0.0681)  & (0.0634)  & (0.0710) & (0.0666)\\
			\hline
			$\varphi_1$ &  16.7480 &  &  &  & $\varphi_2$ & 8.6527    & &\\ 
			&(7.1843)&  &  &   &  &(9.2730) & & &\\
			$c_1$ &  -0.0005 &  &  &  & $c_2$ & 2.0824    & &\\ 
			&(0.0164)&  &  &   &  &(2.5809) & & &\\          
			\bottomrule
		\end{tabular}
	\end{adjustbox}
\end{table}

\begin{figure}[t]
	\centering
	\subfigure[]{\includegraphics[height=4.5cm,width=6.7cm]{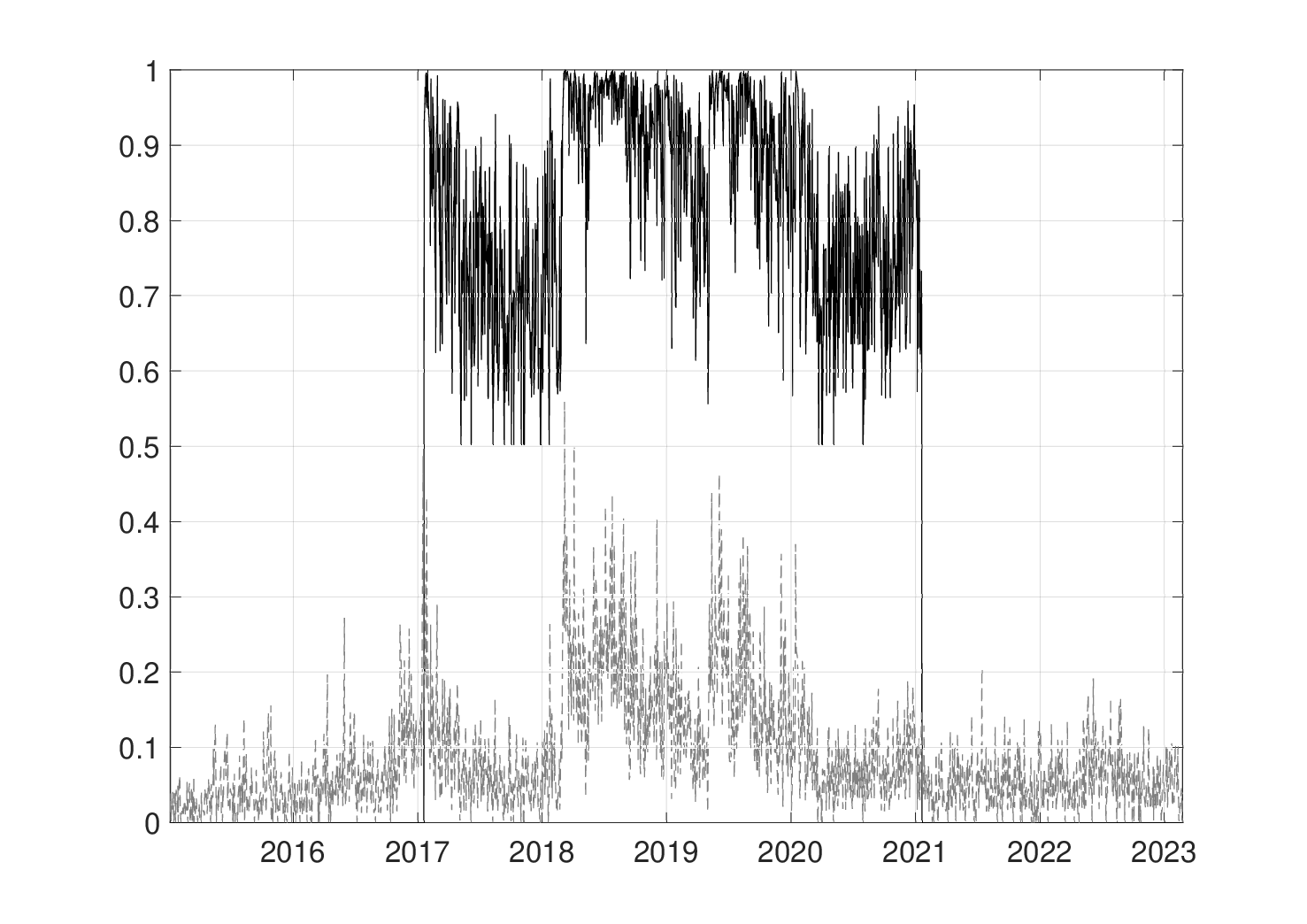}}
	\subfigure[]{\includegraphics[height=4.5cm,width=6.7cm]{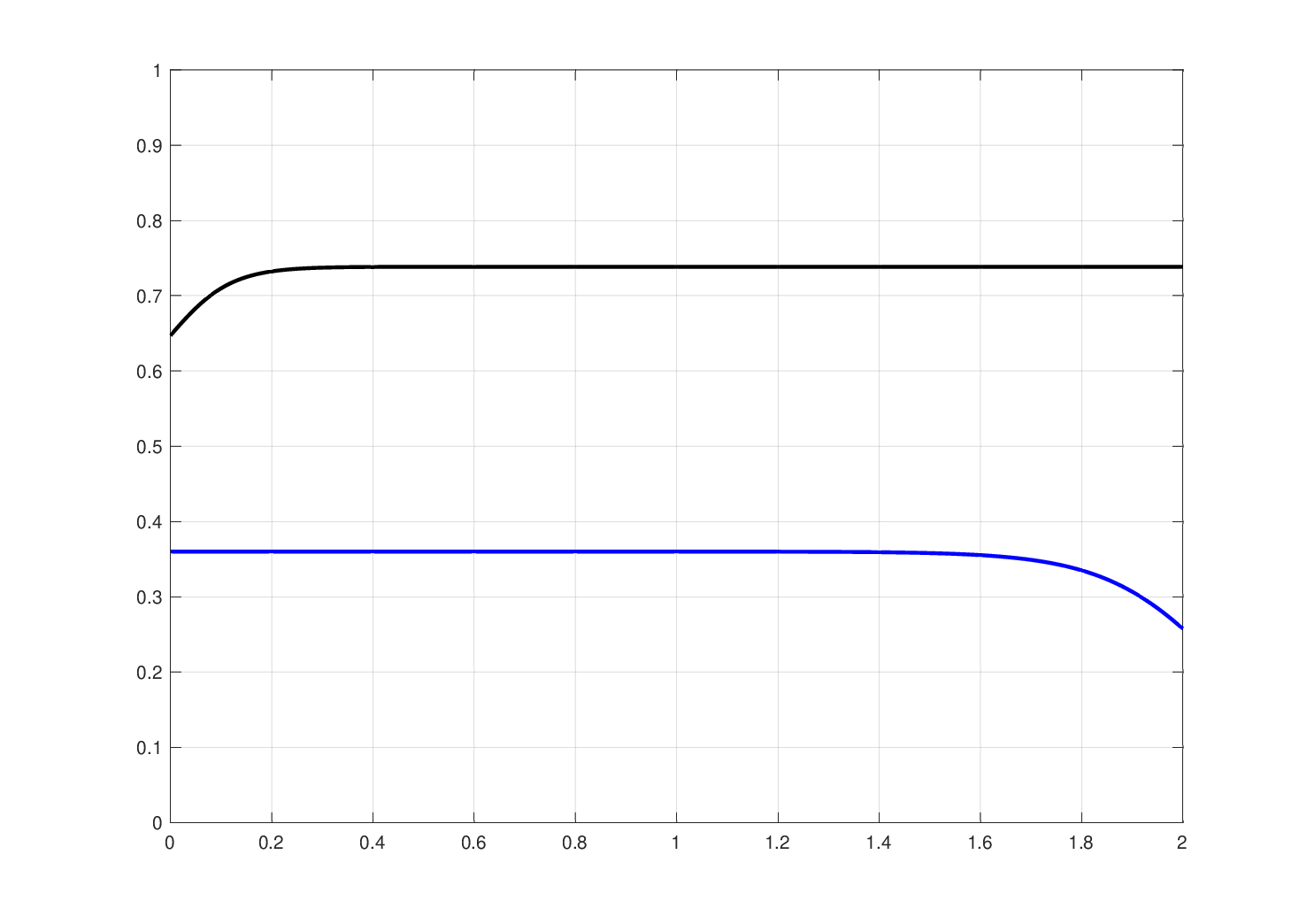}}
	\caption{Panel a) estimated  smooth transition function (black line) and Trade Policy Uncertainty (TPU) index (dashed-gray line) from the STCC-TUPE. Panel b) analytical smooth transition function for the Republican (black line) and Democratic (blue line) administrations. Sample period: January 5, 2015 -- February 24, 2023.}
	\label{fig:est_double_stcc}
\end{figure}

Table \ref{tab:results_double_stcc} presents the estimates. In the smooth transition function, only ${\varphi}_1$ (Republican slope) is significant. During Republican periods, all correlations in $\mathbf{R}_1$ and $\mathbf{R}_2$ are significant, with stronger values in the high-TPU state. Under Democratic administrations ($D_t=0$), $\mathbf{R}_3$ shows no significant correlations: T-Bond relations are close to zero and the Dow Jones–Nasdaq correlation is slightly negative (-0.0525), a result occasionally observed in the literature \citep{Chiang:Yu:Wu:2009}. This may reflect index composition effects or reactions to shocks such as the dot-com bubble or sharp hikes in the Federal Funds Rate.
In contrast, $\mathbf{R}_4$ displays significant correlations, with T-Bond coefficients notably lower than those in $\mathbf{R}_2$. This highlights the role of the presidential dummy, even when ${\varphi}_2$, which represents the  trade uncertainty effect, is not significant. Panel a) of Figure \ref{fig:est_double_stcc} illustrates how the transition function tracks the TPU index: flat during 2015–2017 and 2021–2023, but consistently above 0.5 with high variability during Republican terms. Overall, the STCC-TUPE model shows that bond correlations are shaped by trade policy uncertainty, with stronger effects in high-uncertainty phases, while political cycles further amplify the differences between administrations. This is further supported by panel b) of Figure \ref{fig:est_double_stcc}, which displays the analytical smooth transition functions for the Republican (black) and Democratic (blue) regimes. Notably, the Republican function increases up to a TPU threshold of approximately 0.2 and then remains constant, whereas the Democratic function shows a negative slope for TPU values exceeding 1.6. This suggests that correlations react strongly even to modest increases in trade policy uncertainty under Republican administrations, while Democratic administrations appear less sensitive until TPU levels become relatively high; Moreover, this asymmetry indicates that the political regime does not only shift the level of the transition function but also changes the sensitivity of correlations to TPU

\subsection{Dynamic Conditional Correlation Models}
\label{sec:cdcc}
The models discussed so far do not specify any dynamics in conditional correlations. To address this, we employ the Dynamic Conditional Correlation (DCC) model of \citet{Engle:2002}, where correlations depend on their lags and on lagged products of de-garched returns. We extend the specification by including the TPU index as an exogenous variable and by allowing parameters to vary with the presidential cycle dummy. This follows the VDCC framework of \citet{Bauwens:Otranto:2016}, originally applied with volatility proxies. For the technical details we refer to that paper.

The general model (with correlation targeting) is:

\begin{align}
	\mathbf{R}_t =& \tilde{\mathbf{Q}}_t^{-1}\mathbf{Q}_t\tilde{\mathbf{Q}}_t^{-1}, 
	\quad \tilde{\mathbf{Q}}_t = \mathrm{diag}(\sqrt{Q_{11,t}}, \ldots, \sqrt{Q_{nn,t}})\notag  \\[2mm]
	\mathbf{Q}_t = & \mathbf{Q}^{(1)}_t D_t+\mathbf{Q}^{(2)}_t (1-D_t) \label{eq:dcc_full}
	\\
	\mathbf{Q}^{(d)}_t = &\left[(1-a_d-b_d-\psi_d \bar{x}_{t-1}\right)\bar{\mathbf{R}} + a_d\tilde{\mathbf{Q}}_{t-1}\tilde{\bm{\epsilon}}_{t-1}\tilde{\bm{\epsilon}}_{t-1}^\prime\tilde{\mathbf{Q}}_{t-1}  
	+ b_d\mathbf{Q}_{t-1} + \psi_d x_{t-1} \notag\\
	d=&1,2\notag
\end{align}
Here, $\bar{\mathbf{R}}$ is the unconditional correlation matrix, estimated iteratively from the sample correlation of $\tilde{\mathbf{Q}}_{t}\tilde{\bm{\epsilon}}_{t}$.  The second term, with coefficient $a_d$ ($d=1,2$), depends on the lagged outer product of the de-garched returns, with the correction of \citet{Aielli:2013} to ensure consistency; coefficient $b_d$ captures the autoregressive component, while $\psi_d$ measures the effect of trade uncertainty. With the inclusion of the presidential dummy, the full model in (\ref{eq:dcc_full}) is denoted DCC-TUPE.
The stationarity condition given in \citet{Engle:2002} holds for any set of coefficients with suffix $d=1,2$. Since $x_t$ is non-negative, we expect $\psi_d>0$, which is consistent with the requirement that $\mathbf{Q}_t$ be positive-definite.
By imposing restrictions on the coefficients we obtain alternative specifications:
\begin{itemize}
	\item the classical DCC \citep{Engle:2002}, setting $a_1=a_2$, $b_1=b_2$, and $\psi_1=\psi_2=0$;
	\item the DCC-TUE, i.e. the DCC with a Trade Uncertainty Effect, where $a_1=a_2$, $b_1=b_2$, and $\psi_1=\psi_2$;
	\item the DCC-TUPE$_{\psi}$, imposing $a_1=a_2$ and $b_1=b_2$ but allowing $\psi_1 \neq \psi_2$;
	\item the DCC-PE (Political Effect), where $\psi_1=\psi_2=0$.
\end{itemize}

Estimation results are reported in Table \ref{tab:results_dcc}. With $b_1=0.791$, the DCC shows that correlations adjust slowly, with current estimates heavily influenced by lagged correlations. Correlations also respond to past shocks, as shown by the relatively high $ a_1=0.191$. When the TPU index is included (DCC-TUE, column 2), $ b_1$ increases while $\ a_1$ decreases, and the TPU effect is significant with ${\psi}_1=0.0014$. This suggests that part of the short-run adjustment in correlations is explained by trade uncertainty.

All estimated coefficients remain significant at the 1\% level when we allow for regime-specific parameters (DCC-TUPE, column 3). Under Democratic administrations, lagged correlations exert a stronger effect ($ b_2> b_1$). Conversely, under Republican administrations, both the impact of recent shocks and the effect of the lagged TPU index are stronger, with $\psi_1$ almost double $\psi_2$. In the DCC-TUPE$_\psi$, we allow trade uncertainty to have different effects across administrations. Relative to the DCC-TUE, $ a_1$ and $ b_1$ remain largely unchanged, while ${\psi}_2$ is slightly larger than ${\psi}_1$. This indicates that trade uncertainty affects not only the level but also the dynamics of correlations: when $a$ and $b$ are constrained to be identical across administrations, trade uncertainty has a stronger impact during Democratic periods. Finally, the importance of regime-specific effects is reinforced by the DCC-PE, which shows that correlations are more sensitive to shocks under Republican administrations, but also display lower persistence.

\begin{table}[t]
	\caption{Estimated correlations matrix from the DCC-type models with robust standard errors \citep{White:1980} in parenthesis. Sample period: January 5, 2015 -- February 24, 2023.}\label{tab:results_dcc}
	\begin{adjustbox}{max width=1\linewidth,center}
		\begin{tabular}{lccccc}
			\hline
			\multicolumn{1}{c}{}          &  DCC          & DCC-TUE  & DCC-TUPE & DCC-TUPE$_{\psi}$ & DCC-PE   \\
			\hline
			$a_1$      & 0.1908       & 0.0390   & 0.0454    & 0.0390   & 0.2599   \\
			& (0.0131)     & (0.0000) & (0.0122)  & (0.0005) & (0.0304) \\
			$b_1$       & 0.7907       & 0.9609   & 0.9464    & 0.9590   & 0.7207   \\
			& (0.0128)     & (0.0000) & (0.0132)  & (0.0001) & (0.0215) \\
			$\psi_1$        &              & 0.0014   & 0.0696    & 0.0188   &          \\
			&              & (0.0003) & (0.0391)  & (0.0040) &          \\
			$a_2$      &     &          & 0.0375    &          & 0.1895   \\
			&              &          & (0.0032)  &          & (0.0110) \\
			$b_2$        &              &          & 0.9589    &          & 0.8104   \\
			&              &          & (0.0022)  &          & (0.0105) \\
			$\psi_2$        &              &          & 0.0366    & 0.0207   &          \\
			&              &          & (0.0123)  & (0.0047) &          \\[2mm]
			\bottomrule
		\end{tabular}
	\end{adjustbox}
\end{table}
\begin{figure}[h!]
	\centering
\subfigure{\includegraphics[height=4.5cm,width=6.7cm]{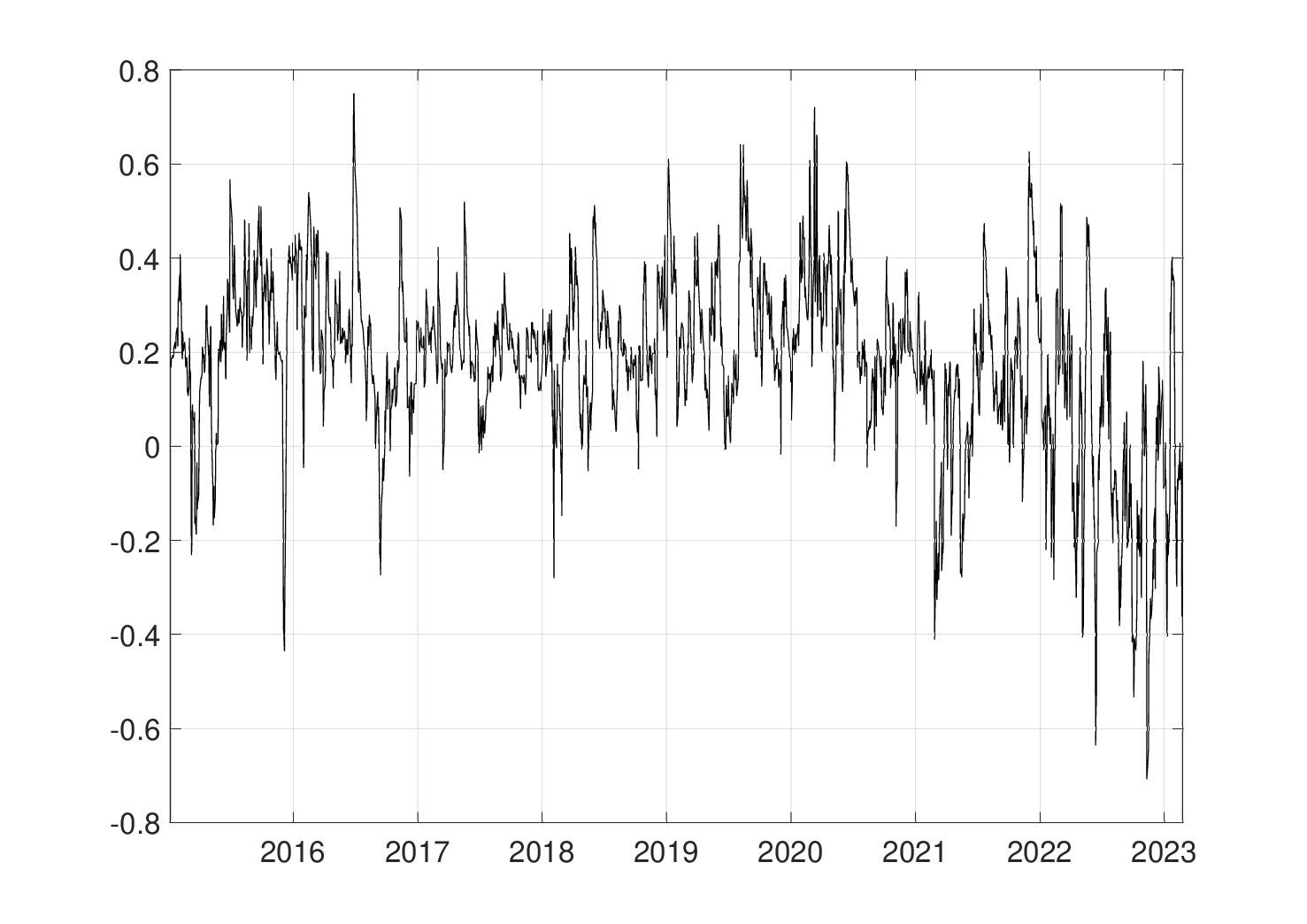}}
\subfigure{\includegraphics[height=4.5cm,width=6.7cm]{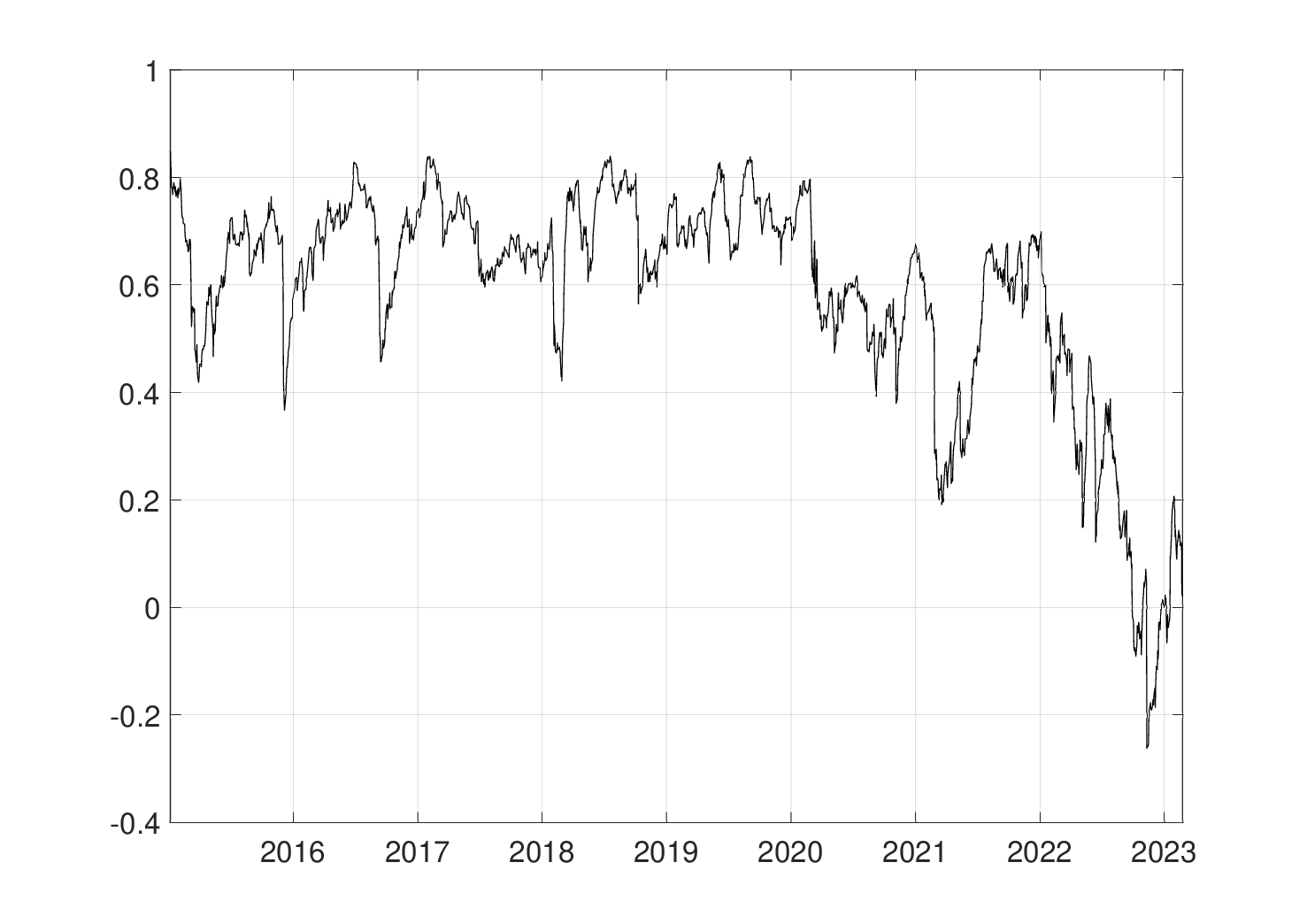}}
	\caption{DCC (a) and DCC-TUPE (b) $S\&P500$-$T-Bond$ estimated correlations.  Sample period: January 5, 2015 -- February 24, 2023.}
	\label{fig:est_dcc}
\end{figure}
Figure \ref{fig:est_dcc} illustrates the estimated S\&P500–T-Bond correlations with standard DCC (panel a) and DCC-TUPE (panel b). In the standard DCC, correlations appear more erratic due to the relatively low $ b_1$, whereas in the DCC-TUPE persistence emerges clearly. Moreover, the DCC-TUPE highlights a distinct political effect: during the most recent Democratic administration, starting in 2022, correlations follow a marked negative trend, coinciding with lower levels of trade policy uncertainty.

\section{Evaluating Competitive Models}
\label{sec:comparison}
In this section, we compare the nine estimated models both in-sample (Section \ref{sec:insample}) and out-of-sample (Section \ref{sec:forecast}). The in-sample analysis evaluates model fit and employs statistical tests to identify the best-fitting specification, while also verifying whether correlations are indeed time-varying. Conversely, the out-of-sample exercise assesses forecasting ability through the Model Confidence Set (MCS) approach \citep{Hansen:Lunde:Nason:2011}.
\subsection{In-sample Model Comparison}
\label{sec:insample}
The in-sample performance is first evaluated using the standard information criteria, AIC and BIC. Results, reported in Table \ref{tab:aic_bic} (panel a), best model in bold), clearly highlight the advantage of time-varying correlation structures. The DCC-TUPE$_{\psi}$ emerges as the preferred model under both AIC and BIC. These criteria favor a specification where correlations evolve dynamically, driven by both the TPU index (as an exogenous variable) and the presidential cycle (capturing regime differences). Notably, the DCC family is also the most parsimonious (see the last row of panel a), reporting the number of parameters), since CCC and STCC models require direct estimation of all correlation parameters, while the DCC specification derives the unconditional correlation matrix iteratively through standardized de-garched returns and correlation targeting (see Eq. \ref{eq:dcc_full}). Despite higher log-likelihoods for CCC and STCC models, the information criteria consistently favor DCC specifications with TPU effects, reinforcing the importance of this factor in shaping correlation dynamics.

The role of TPU and the presidential dummy within the DCC family is evaluated through t-tests on the corresponding coefficients (Table \ref{tab:results_dcc}). For other families, significance is tested using a Likelihood Ratio (LR) test to assess the presidential dummy in CCC models (CCC vs. CCC-PE), and the Lagrange Multiplier (LM) test of \citet{Silvennoinen:Terasvirta:2009} to evaluate whether a TPU-driven smooth transition is necessary (STCC-TUE vs. CCC and STCC-TUPE vs. CCC-PE).\footnote{The standard LR test cannot be applied here because of the nuisance parameter problem: the location parameter $c$ appears only under the null, yielding an unknown asymptotic distribution of the test statistic.} In all cases, the null hypothesis is rejected, confirming that correlations are time-varying and that both TPU and the presidential dummy provide valuable explanatory power.

Panel c) of Table \ref{tab:aic_bic} focuses on stock–bond correlations. We employ a Wald statistic ($\mathcal{W}$) to test the equality of stock–bond correlations in matrices $\mathbf{R}_1$ and $\mathbf{R}_2$ for the CCC-PE (Eq. \ref{eq:ccc-pe}) and STCC-TUE (Eq. \ref{eq:stcc}) models, and across $\mathbf{R}_1$ vs. $\mathbf{R}_2$ and $\mathbf{R}_3$ vs. $\mathbf{R}_4$ for the STCC-TUPE (Eq. \ref{eq:stcc_tupe}). The null hypothesis is rejected in all cases, indicating that stock–bond correlations differ significantly across alternative scenarios.

\begin{table}[t]
	\caption{a) AIC and BIC information criteria for the estimated models (best model in bold); b) Lagrange Multiplier (LM) and Likelihood Ratio (LR) tests for the hypothesis of constant correlations; c) Wald ($\mathcal{W}$)  test for the for the hypothesis of constant stock-bond correlations and; d) LR test for nested models. Critical values at a 5$\%$ significance level  are reported in square brackets.  Sample period: January 5, 2015 -- February 24, 2023.}\label{tab:aic_bic}
	\begin{adjustbox}{max width=1\linewidth,center}
		
		\begin{tabular}{lccccccccc}
			\hline
			&\multicolumn{9}{c}{\textbf{a) Information Criteria}} \\ 
			& CCC     & CCC-PE  & STCC-TUE & STCC-TUPE & DCC     & DCC-TUE & DCC-TUPE & DCC-TUPE$_\psi$ & DCC-PE  \\
			AIC & -3.4528 & -3.5180 & -3.4800  & -3.5446   & -3.2670 & -3.7584 & -3.7630  & \textbf{-3.7631}         & -3.3270 \\
			BIC & -3.4306 & -3.4735 & -3.4312  & -3.4469   & -3.2626 & -3.7517 & -3.7496  & \textbf{-3.7542}         & -3.3181\\
			\# of parameters & 10 & 20 & 22 & 44 & 2 & 3 & 6 & 4 & 4 \\
		\end{tabular}
	\end{adjustbox}
	\begin{adjustbox}{max width=1\linewidth,center}
		\begin{tabular}{lccc}
			\hline
			& \multicolumn{3}{c}{\textbf{b) Testing $H_0$ of constant correlations}}  \\
			& LR& \multicolumn{2}{c}{LM}  \\[2mm]
			& CCC-PE vs CCC &STCC-TUE vs CCC &  STCC-TUPE vs CCC-PE \\
			& 192.6014 & 88.5000             & 22452240.0000        \\
			&  [18.3070]& [21.0261]            & \multicolumn{1}{c}{[36.4150]}        \\
			\bottomrule
			& \multicolumn{3}{c}{\textbf{c) Testing $H_0$ of constant stock-bond correlations}}  \\
			& CCC-PE       & STCC-TUE       & STCC-TUPE      \\
			$\mathcal{W}$ &  131.9417           & 273.5868      & 509.8305  \\ 
			&[9.4877]           & [9.4877]      & [15.5073] \\
		\end{tabular}
	\end{adjustbox}
	
	\begin{adjustbox}{max width=1\linewidth,center}
		\begin{tabular}{lcccc}
			\bottomrule
			& \multicolumn{3}{c}{\textbf{d) Testing nested models}}\\
			& STCC-TUPE vs STCC-TUE       & DCC-TUPE vs DCC    & DCC-TUPE vs DCC-TUE        \\
			LR & 215.0006                    & 1321.2946          & 18.1816                    \\
			& [33.9244]                    & [9.4877]             & [7.8147]                     \\[3mm]
			& DCC-TUPE vs DCC-TUPE$_\psi$ & DCC-TUPE vs DCC-PE & DCC-TUPE$_\psi$ vs DCC-TUE \\
			LR & 3.6456                      & 1158.3964          & 14.5360                    \\
			& [5.9915]                      & [5.9915]             & [3.8415]                     \\[3mm]
			& DCC-TUE vs DCC              & DCC-PE vs DCC      & DCC-TUPE$_\psi$ vs DCC     \\
			LR & 1303.1130                   & 162.8982           & 1317.6490                  \\
			& [3.8415]                      & [5.9915]             & [5.9915] \\
			\bottomrule
		\end{tabular}
	\end{adjustbox}
\end{table}

Finally, panel d) reports LR tests for nested specifications, assessing whether additional parameters significantly improve the model’s fit. In nearly all cases, the LR statistics exceed the 5\% chi-squared critical values, showing that both STCC and DCC models provide a superior fit relative to their restricted versions. The only exception is DCC-TUPE versus DCC-TUPE$_\psi$, where the LR statistic (3.646) falls short of the 5\% threshold (5.992). This suggests that the GARCH-like correlation dynamics are homogeneous across scenarios shaped by trade uncertainty and presidential regimes, while the exogenous TPU effect remains heterogeneous, confirming its key role in driving correlation dynamics.

\subsection{Out-of-Sample Forecasting Comparison}
\label{sec:forecast}

The out-of-sample analysis relies on a rolling window scheme with a fixed estimation sample of 2,049 observations and 60 out-of-sample forecasts per iteration. The window is rolled forward by 60 observations each time, yielding 10 estimation–forecasting periods and a total of 600 out-of-sample forecasts. Model comparison is conducted using the MCS procedure \citep{Hansen:Lunde:Nason:2011}, which identifies, for a given significance level and loss function, the set of best models with similar forecasting performance.
We consider two loss functions. First, the $QLike$ loss, which is consistent according to \citet{Patton:2011}, defined as:
\begin{equation}
	Qlike=\frac{1}{T_h}\sum_{\tau=1}^{T_h}\left(ln|\mathbf{\hat{H}_\tau}|+trace(\mathbf{\hat{H}_\tau}^{-1}\mathbf{C}_\tau\right)\label{eq:qlike},
\end{equation}
where $\hat{\mathbf{H}}_\tau$ denotes the forecasted covariance matrix, $\mathbf{C}_\tau=\bm{r}_\tau'\bm{r}_\tau$ is the realized covariance proxy, and $T_h$ is the number of forecasts.
Second, we consider an economic loss function based on the Global Minimum Variance ($GMV$) portfolio \citep{englecolacito2006}, constructed from forecasted covariances as:
\begin{equation}
	GMV=\frac{1}{T_h}\sum_{\tau=1}^{T_h}\left(\hat{\bm{v}}^\prime_\tau \mathbf{\hat{H}_\tau}\hat{\bm{v}}_\tau\right)\label{eq:gmv},
\end{equation}
with $\hat{\bm{v}}_\tau=\sqrt{n}\hat{\mathbf{H}}_\tau^{-1}\mathbf{j}_n/\big(\mathbf{j}'_n \hat{\mathbf{H}}_\tau^{-1}\mathbf{j}_n\big)$, where $\mathbf{j}_n$ is a vector of ones.

\begin{table}[t]
	\caption{Model Confidence Set with  $Qlike$ and $GMV$ as loss functions. The first row represents the first model removed, down to the best performing model in the last row.Set of superior models and associated p-values in bold.}\label{tab:mcs}
	\begin{adjustbox}{max width=1\linewidth,center}
		\begin{tabular}{lclcclclc}
			\hline
			\multicolumn{4}{c}{$T_R$}                                                                                                 &                         & \multicolumn{4}{c}{$T_{SQ}$} \\
			\multicolumn{2}{c}{$Qlike$}                                              & \multicolumn{2}{c}{$GMV$}                                       &                       & \multicolumn{2}{c}{$Qlike$}                                         & \multicolumn{2}{c}{$GMV$}                                       \\\hline
			Model                                  & p-value                       & Model                         & p-value                       &    \multicolumn{1}{c|}{}                     & Model                           & p-value                         & Model                         & p-value                       \\[2mm]
			DCC       & 0.0000 & STCC-TUE        & 0.0000 &  & DCC       & 0.0000 & STCC-TUE        & 0.0000 \\
			DCC-PE    & 0.0000 & DCC-TUPE        & 0.0000 &  & DCC-PE    & 0.0000 & DCC-TUPE        & 0.0000 \\
			CCC       & 0.0000 & DCC-TUPE$_\psi$ & 0.0000 &  & CCC       & 0.0000 & DCC-TUPE$_\psi$ & 0.0000 \\
			STCC-TUPE & 0.0000 & DCC-TUE         & 0.0000 &  & STCC-TUPE & 0.0000 & DCC-TUE         & 0.0000 \\
			CCC-PE    & 0.0000 & STCC-TUPE       & 0.0000 &  & CCC-PE    & 0.0000 & STCC-TUPE       & 0.0006 \\
			STCC-TUE  & 0.0000 & CCC-PE          & 0.0000 &  & STCC-TUE  & 0.0000 & CCC-PE          & 0.0006 \\
			DCC-TUE   & 0.0009 & CCC             & 0.0000 &  & DCC-TUE   & 0.0057 & CCC             & 0.0006\\
			\textbf{DCC-TUPE}        & \textbf{ 0.2309} & \textbf{DCC}    & \textbf{0.926} &  & \textbf{DCC-TUPE}        & \textbf{0.2292} & \textbf{DCC}    & \textbf{0.9274} \\
			\textbf{DCC-TUPE$_\psi$} &        & \textbf{DCC-PE} &       &  & \textbf{DCC-TUPE$_\psi$} &        & \textbf{DCC-PE} &  \\
			\bottomrule
		\end{tabular}
	\end{adjustbox}
\end{table}

Table \ref{tab:mcs} reports the MCS p-values, highlighting in bold the models included in the superior set. Results are consistent across both test statistics, the Range ($T_R$) and Semi-Quadratic ($T_{SQ}$).\footnote{See \citet{Hansen:Lunde:Nason:2011} for details about these statistics.} First, the CCC model is systematically excluded, confirming that constant correlations are too restrictive even out-of-sample. Second, DCC specifications dominate STCC alternatives. Under the $QLike$ loss, accounting for both trade uncertainty and political effects is crucial, as the DCC-TUPE and DCC-TUPE$_\psi$ deliver the strongest forecasting performance. Under the $GMV$ loss, the DCC with political effects emerges as the best model, although the simpler DCC performs comparably well.

\section{Concluding Remarks}
\label{sec:conclusion}
We argue that the correlation between U.S. stocks and bonds is influenced by trade uncertainty and the political cycle. To investigate this, we extend the most widely used conditional correlation models (CCC, STCC, DCC) to incorporate two proxies for these effects: the TPU index to capture trade uncertainty, and a presidential dummy to distinguish Republican from Democratic terms. Our multivariate analysis considers four major U.S. stock market indices and 10-year Treasury bonds.

We proceed by gradually increasing model flexibility, starting from the CCC benchmark with constant correlations and no external effects, up to the DCC-TUPE, which combines a GARCH-type dynamic structure with both sources of variation. By presenting estimation results alongside each model, we show how fit and interpretability improve as flexibility increases. Out-of-sample analysis further confirms the superior performance of DCC models, particularly those that account for both effects.
Our findings indicate that stock–bond correlations rise in periods of high trade uncertainty and during Republican administrations, which often introduced trade shocks such as new tariffs. Correlations shift from strongly positive values in times of high uncertainty -- especially under Republican administrations -- to zero or even negative values during periods of low uncertainty. In contrast, correlations among stock indices remain persistently high and largely unaffected by these factors.
These results highlight the importance of monitoring stock–bond correlations. High correlations imply similar behavior across the two asset classes, undermining the role of bonds as portfolio stabilizers.

In conclusion, the recent development of multivariate models for conditional correlations opens avenues for further research. For instance, the VDCC class of models proposed by \citet{Bauwens:Otranto:2016}, originally designed to capture the impact of volatility on financial correlations, could be extended to this framework. In this sense, our analysis points to a promising line of inquiry that may be pursued with alternative methodological tools. Finally, by focusing on individual stocks rather than equity indices, it would be possible to further investigate the impact of TPU at the firm level, with the possibility of detecting asymmetric and heterogeneous responses.

\subsection*{Conflicts of interest}
The authors declare that they have no competing interests.

\subsection*{Funding}
Financial support is acknowledged by Otranto: Italian  PRIN 2022 grant (20223725WE) ``Methodological and computational issues in large-scale time series models for economics and finance". The views and opinions expressed are solely those of the authors and do not necessarily reflect those of the EU, nor can the EU be held responsible for them. 

\subsection*{Data availability}
The data underlying this article are available upon request

\subsection*{Author contributions statement}
\textbf{D.L:} Writing -- Original Draft, Methodology, Investigation, Conceptualization, Formal analysis, Data Curation, Validation.
 \textbf{E.O:} Writing -- Review \& Editing, Supervision, Methodology, Conceptualization, Validation, Formal analysis, Funding acquisition. 

\bibliographystyle{oup-abbrvnat}
\bibliography{biblio}

\begin{thebibliography}{30}
\providecommand{\natexlab}[1]{#1}
\providecommand{\url}[1]{\texttt{#1}}
\expandafter\ifx\csname urlstyle\endcsname\relax
  \providecommand{\doi}[1]{doi: #1}\else
  \providecommand{\doi}{doi: \begingroup \urlstyle{rm}\Url}\fi

\bibitem[Aielli(2013)]{Aielli:2013}
G.~P. Aielli.
\newblock Dynamic conditional correlation: on properties and estimation.
\newblock \emph{Journal of Business \& Economic Statistics}, 31\penalty0
  (3):\penalty0 282--299, 2013.

\bibitem[Andersson et~al.(2008)Andersson, Krylova, and
  V{\"a}h{\"a}maa]{Andersson:Krylova:Vahamaa:2008}
M.~Andersson, E.~Krylova, and S.~V{\"a}h{\"a}maa.
\newblock Why does the correlation between stock and bond returns vary over
  time?
\newblock \emph{Applied Financial Economics}, 18\penalty0 (2):\penalty0
  139--151, 2008.

\bibitem[Baele et~al.(2010)Baele, Bekaert, and
  Inghelbrecht]{Baele:Bekaert:Inghelbrecht:2010}
L.~Baele, G.~Bekaert, and K.~Inghelbrecht.
\newblock The determinants of stock and bond return comovements.
\newblock \emph{The Review of Financial Studies}, 23\penalty0 (6):\penalty0
  2374--2428, 2010.

\bibitem[Baker et~al.(2016)Baker, Bloom, and Davis]{Baker:Bloom:Davis:2016}
S.~R. Baker, N.~Bloom, and S.~J. Davis.
\newblock Measuring economic policy uncertainty.
\newblock \emph{The Quarterly Journal of Economics}, 131\penalty0 (4):\penalty0
  1593--1636, 2016.

\bibitem[Bauwens and Otranto(2016)]{Bauwens:Otranto:2016}
L.~Bauwens and E.~Otranto.
\newblock Modeling the dependence of conditional correlations on market
  volatility.
\newblock \emph{Journal of Business \& Economic Statistics}, 34\penalty0
  (2):\penalty0 254--268, 2016.

\bibitem[Bauwens et~al.(2006)Bauwens, Laurent, and
  Rombouts]{Bauwens:Laurent:Rombouts:2006}
L.~Bauwens, S.~Laurent, and J.~V.~K. Rombouts.
\newblock Multivariate {GARCH} models: a survey.
\newblock \emph{Journal of Applied Econometrics}, 21:\penalty0 79--109, 2006.

\bibitem[Bekaert et~al.(2010)Bekaert, Engstrom, and
  Grenadier]{Bekaert:Engstrom:Grenadier:2010}
G.~Bekaert, E.~Engstrom, and S.~R. Grenadier.
\newblock Stock and bond returns with moody investors.
\newblock \emph{Journal of Empirical Finance}, 17\penalty0 (5):\penalty0
  867--894, 2010.

\bibitem[Bloom(2009)]{Bloom:2009}
N.~Bloom.
\newblock The impact of uncertainty shocks.
\newblock \emph{Econometrica}, 77\penalty0 (3):\penalty0 623--685, 2009.

\bibitem[Bollerslev(1990)]{Bollerslev:1990}
T.~Bollerslev.
\newblock Modelling the coherence in short-run nominal exchange rates: a
  multivariate generalized arch model.
\newblock \emph{The Review of Economics and Statistics}, 72\penalty0
  (3):\penalty0 498--505, 1990.

\bibitem[Bollerslev et~al.(1988)Bollerslev, Engle, and
  Wooldridge]{Bollerslev:Engle:Wooldridge:1988}
T.~Bollerslev, R.~F. Engle, and J.~M. Wooldridge.
\newblock A capital asset pricing model with time-varying covariances.
\newblock \emph{Journal of Political Economy}, 96:\penalty0 116--131, 1988.

\bibitem[Brogaard and Detzel(2015)]{Brogaard:Detzel:2015}
J.~Brogaard and A.~Detzel.
\newblock The asset-pricing implications of government economic policy
  uncertainty.
\newblock \emph{Management Science}, 61\penalty0 (1):\penalty0 3--18, 2015.

\bibitem[Caldara et~al.(2020)Caldara, Iacoviello, Molligo, Prestipino, and
  Raffo]{Caldara:Iacoviello:Molligo:Prestipino:Raffo:2020}
D.~Caldara, M.~Iacoviello, P.~Molligo, A.~Prestipino, and A.~Raffo.
\newblock The economic effects of trade policy uncertainty.
\newblock \emph{Journal of Monetary Economics}, 109:\penalty0 38--59, 2020.

\bibitem[Chiang et~al.(2009)Chiang, Yu, and Wu]{Chiang:Yu:Wu:2009}
T.~C. Chiang, H.-C. Yu, and M.-C. Wu.
\newblock Statistical properties, dynamic conditional correlation and scaling
  analysis: Evidence from dow jones and nasdaq high-frequency data.
\newblock \emph{Physica A: Statistical Mechanics and its Applications},
  388\penalty0 (8):\penalty0 1555--1570, 2009.

\bibitem[Connolly et~al.(2005)Connolly, Stivers, and
  Sun]{Connolly:Stivers:Sun:2005}
R.~Connolly, C.~Stivers, and L.~Sun.
\newblock Stock market uncertainty and the stock-bond return relation.
\newblock \emph{Journal of Financial and Quantitative Analysis}, 40\penalty0
  (1):\penalty0 161--194, 2005.

\bibitem[Demirer and Gupta(2018)]{Demirer:Gupta:2018}
R.~Demirer and R.~Gupta.
\newblock Presidential cycles and time-varying bond--stock market correlations:
  Evidence from more than two centuries of data.
\newblock \emph{Economics Letters}, 167:\penalty0 36--39, 2018.

\bibitem[Engle and Colacito(2006)]{englecolacito2006}
R.~Engle and R.~Colacito.
\newblock Testing and valuing dynamic correlations for asset allocation.
\newblock \emph{Journal of Business \& Economic Statistics}, 24\penalty0
  (2):\penalty0 238--253, 2006.

\bibitem[Engle and Kroner(1995)]{Engle:Kroner:1995}
R.~Engle and F.~K. Kroner.
\newblock Multivariate simultaneous generalized arch.
\newblock \emph{Econometric Theory}, 11:\penalty0 122--150, 1995.

\bibitem[Engle(2002)]{Engle:2002}
R.~F. Engle.
\newblock New frontiers for {ARCH} models.
\newblock \emph{Journal of Applied Econometrics}, 17:\penalty0 425--446, 2002.

\bibitem[Fang et~al.(2017)Fang, Yu, and Li]{Fang:Yu:Li:2017}
L.~Fang, H.~Yu, and L.~Li.
\newblock The effect of economic policy uncertainty on the long-term
  correlation between {US} stock and bond markets.
\newblock \emph{Economic Modelling}, 66:\penalty0 139--145, 2017.

\bibitem[Fang et~al.(2018)Fang, Chen, Yu, and Xiong]{Fang:Chen:Yu:Xiong:2018}
L.~Fang, B.~Chen, H.~Yu, and C.~Xiong.
\newblock The effect of economic policy uncertainty on the long-run correlation
  between crude oil and the {US} stock markets.
\newblock \emph{Finance Research Letters}, 24:\penalty0 56--63, 2018.

\bibitem[Glosten et~al.(1993)Glosten, Jagannanthan, and
  Runkle]{Glosten:Jaganathan:Runkle:1993}
L.~R. Glosten, R.~Jagannanthan, and D.~E. Runkle.
\newblock On the relation between the expected value and the volatility of the
  nominal excess return on stocks.
\newblock \emph{The Journal of Finance}, 48\penalty0 (5):\penalty0 1779--1801,
  1993.

\bibitem[Hansen et~al.(2011)Hansen, Lunde, and Nason]{Hansen:Lunde:Nason:2011}
P.~R. Hansen, A.~Lunde, and J.~M. Nason.
\newblock The model confidence set.
\newblock \emph{Econometrica}, 79\penalty0 (2):\penalty0 453--497, 2011.

\bibitem[Pastor and Veronesi(2012)]{Pastor:Veronesi:2012}
L.~Pastor and P.~Veronesi.
\newblock Uncertainty about government policy and stock prices.
\newblock \emph{The Journal of Finance}, 67\penalty0 (4):\penalty0 1219--1264,
  2012.

\bibitem[Patton(2011)]{Patton:2011}
A.~J. Patton.
\newblock Volatility forecast comparison using imperfect volatility proxies.
\newblock \emph{Journal of Econometrics}, 160\penalty0 (1):\penalty0 246 --
  256, 2011.
\newblock ISSN 0304-4076.

\bibitem[Silvennoinen and Ter{\"a}svirta(2009)]{Silvennoinen:Terasvirta:2009}
A.~Silvennoinen and T.~Ter{\"a}svirta.
\newblock Modeling multivariate autoregressive conditional heteroskedasticity
  with the double smooth transition conditional correlation garch model.
\newblock \emph{Journal of Financial Econometrics}, 7\penalty0 (4):\penalty0
  373--411, 2009.

\bibitem[Silvennoinen and Ter{\"a}svirta(2015)]{Silvennoinen:Terasvirta:2015}
A.~Silvennoinen and T.~Ter{\"a}svirta.
\newblock Modeling conditional correlations of asset returns: A smooth
  transition approach.
\newblock \emph{Econometric Reviews}, 34\penalty0 (1-2):\penalty0 174--197,
  2015.

\bibitem[White(1980)]{White:1980}
H.~White.
\newblock A heteroskedasticity-consistent covariance matrix estimator and a
  direct test for heteroskedasticity.
\newblock \emph{Econometrica}, 48\penalty0 (4):\penalty0 817--38, 1980.

\bibitem[Yilmazkuday(2023)]{Yilmazkuday:2023}
H.~Yilmazkuday.
\newblock Covid-19 effects on the {S\&P} 500 index.
\newblock \emph{Applied Economics Letters}, 30\penalty0 (1):\penalty0 7--13,
  2023.

\bibitem[Yilmazkuday(2024)]{Yilmazkuday:2024}
H.~Yilmazkuday.
\newblock Geopolitical risk and stock prices.
\newblock \emph{European Journal of Political Economy}, 83:\penalty0 102553,
  2024.

\bibitem[Yilmazkuday(2025)]{Yilmazkuday:2025}
H.~Yilmazkuday.
\newblock {US} tariffs and stock prices.
\newblock \emph{Finance Research Letters}, 83:\penalty0 107708, 2025.

\end{thebibliography}

\end{document}